\documentclass[sigconf]{acmart}
\makeatletter
\renewcommand\@formatdoi[1]{\ignorespaces}
\makeatother

\setcopyright{rightsretained}
\copyrightyear{2021}

\acmConference[SpaceCHI '21]{SpaceCHI '21: Making Waves, Combining Strengths}{May 14, 2021}{The CHI Workshop on Human-Computer Interaction for Space Exploration}
\acmBooktitle{}
\acmPrice{}
\acmISBN{}

\usepackage{subcaption}
\AtBeginDocument{\AtBeginShipoutNext{\AtBeginShipoutDiscard}}

\begin{document}

\title{PIXLISE-C: Data Analysis for Mineral Identification}{PIXLISE-C: Exploring The Data Analysis Needs of NASA Scientists for Mineral Identification}

\author{Connie Ye}
\authornote{All authors contributed equally to this research.}
\email{constany@andrew.cmu.edu}
\orcid{1234-5678-9012}
\author{Lukas Hermann}
\authornotemark[1]
\email{lhermann@andrew.cmu.edu}
\author{Nur Yildirim}
\email{nyildiri@andrew.cmu.edu}
\authornotemark[1]
\author{Shravya Bhat}
\email{shravyab@andrew.cmu.edu}
\authornotemark[1]
\affiliation{%
  \institution{Carnegie Mellon University}
  \streetaddress{5000 Forbes Avenue}
  \city{Pittsburgh}
  \state{Pennsylvania}
  \country{USA}
  \postcode{43017-6221}
}

\author{Dominik Moritz}
\affiliation{%
  \institution{Carnegie Mellon University}
  \streetaddress{5000 Forbes Avenue}
  \city{Pittsburgh}
  \state{Pennsylvania}
  \country{USA}}
\email{domoritz@cmu.edu}

\author{Scott Davidoff}
\affiliation{%
  \institution{Jet Propulsion Laboratory \\ California Institute of Technology}
  \streetaddress{1 Th{\o}rv{\"a}ld Circle}
  \city{Pasadena}
  \state{California}
  \country{USA}}
\email{scott.davidoff@jpl.nasa.gov}

\renewcommand{\shortauthors}{Ye and Hermann and Yildirim and Bhat et all.}

\begin{abstract}
NASA JPL scientists working on the micro x-ray fluorescence (microXRF) spectroscopy data collected from Mars surface perform data analysis to look for signs of past microbial life on Mars. Their data analysis workflow mainly involves identifying mineral compounds through the element abundance in spatially distributed data points. Working with the NASA JPL team, we identified pain points and needs to further develop their existing data visualization and analysis tool. Specifically, the team desired improvements for the process of creating and interpreting mineral composition groups. To address this problem, we developed an interactive tool that enables scientists to (1) cluster the data using either manual lasso-tool selection or through various machine learning clustering algorithms, and (2) compare the clusters and individual data points to make informed decisions about mineral compositions. Our preliminary tool supports a hybrid data analysis workflow where the user can manually refine the machine-generated clusters. 
\end{abstract}

\ccsdesc[100]{Applied computing~Chemistry}

\keywords{Data Analysis, PIXL, Mars 2020, D3, Data Visualization}


\maketitle

\section{Introduction}
In collaboration with NASA Jet Propulsion Laboratory Human-Centered Design Group, we developed a tool to support the exploratory analysis of astrobiological data collected via Mars 2020 Perseverance rover. One objective of NASA scientists is to to search for evidence of past microbial life on other bodies in our solar system, with a current focus on Mars. Geologists on Earth have examined the fine scale geochemical and morphological structures in rocks to identify micron-scale evidence of biological mediation -- in essence, fossils that strongly suggest biogenic causes \cite{Banfield2001, Allwood9548}.

The search for life on other planets follows a similar process, today with scientific instruments mounted on spacecrafts. NASA’s Perseverance Rover carries multiple astrobiology instruments\cite{Farley2020}, including a micro x-ray fluorescence (microXRF) spectrometer, called the Planetary Instrument for X-ray Lithochemistry (PIXL)\cite{Allwood2020}. PIXL will be used by scientists to examine the concentration and the spatial distribution of elements. This data will be used to determine the mineral composition of the rocks and to look for evidence of fossil activity mediated by microbes \cite{PIXL4Scientists}.

An initial data visualisation tool prototype, PIXLISE, is already in place to support scientists in this process \cite{PixlDatavis, Allwood2020}. Our preliminary discussions with the project team, therefore, focused on identifying the tasks and information needs of the scientists that might not be supported by the current app \cite{PIXLISECOSPAR2021}. In PIXLISE \cite{PIXLISEwww}, the data analysis tasks are focused around a map interface, as the collected spectroscopy data are spatially localized within a sample area and co-registered within the associated image\cite{Allwood2020}. Our observations of the science team conclude that the data analysis workflow has two major stages: (1) identifying peaks in spectra and labeling them as elements whose abundance is then computed; and (2) creating and interpreting mineral composition groups of the underlying rock based on spatial covariance of previously computed elemental abundance. Through a collaborative problem-defining process \cite{sanders2008co}, we decided to focus on mineral identification, as the science team guided us to understand that mineral identification is a problem that affects more of the science team, and whose technical support in PIXLISE was the least well developed.

In this paper, we first discuss the existing application (PIXLISE) and the workflow with a focus on information needs. Next, we introduce the improved tool we propose (PIXLISE-C), and finally we discuss the design implications for future tool development.

\section{Related Work}
Across decades, instruments studying Martian geology aboard spacecraft from Odyssey \cite{Saunders2001}, to Pathfinder \cite{Rieder1771}, and Mars Science Laboratory \cite{Wiens2012} have been focusing on elemental composition data to describe the mineral species. These analyses have investigated geochemical data to elaborate the bulk composition of specific rocks. These interpretations often examine co-varying elemental maps to infer minerals and their abundances \cite{VanBommel2016, Ruff2016, MCLENNAN200595}.

The PIXL instrument on board the Perseverance rover brings unprecedented resolution to this type of mineralogical investigation. Scans from PIXL may contain as many as 15,000 spatially resolved sample points, with samples on the scale of 100 microns \cite{Allwood2020}. Mineral analyses using these data will be based several orders of magnitude richer than those from previous instruments, opening many opportunities to better support the process of scientific analysis of large numbers of spatially resolved spectroscopy data.

In the field of data visualization, how large amounts of data can be visualized is an ongoing exploration; in his 1996 paper, Shneiderman performs a taxonomy of different methods for visualization and transformation, such as filtering, zooming, and more \cite{Shneiderman}. Reacting to the emergence of artificial intelligence, Horvitz suggests that human-AI interfaces can be designed to support collaboration by pairing automated services with direct manipulation \cite{Horvitz}. 

\section{Baseline dataset, PIXLISE App and the Data Analysis Workflow}

The PIXL data set is a collection of spatially localized spectroscopy data as the instrument on Perseverance rover passes over the sampling area \cite{Allwood2020}. The microXRF data is sensed by the instrument and stored as 1024 ordinal channels, each, which after energy calibration, encodes a count of the energy of the x-rays sensed by the instrument's 2 detectors at that particular channel (in kilo electron-volts). The dataset is organized into a CSV file that contains information on points such as coordinates (X, Y, Z) and 4000 spectral channels.

\begin{figure}[h]
    \centering
    \includegraphics[width=\linewidth]{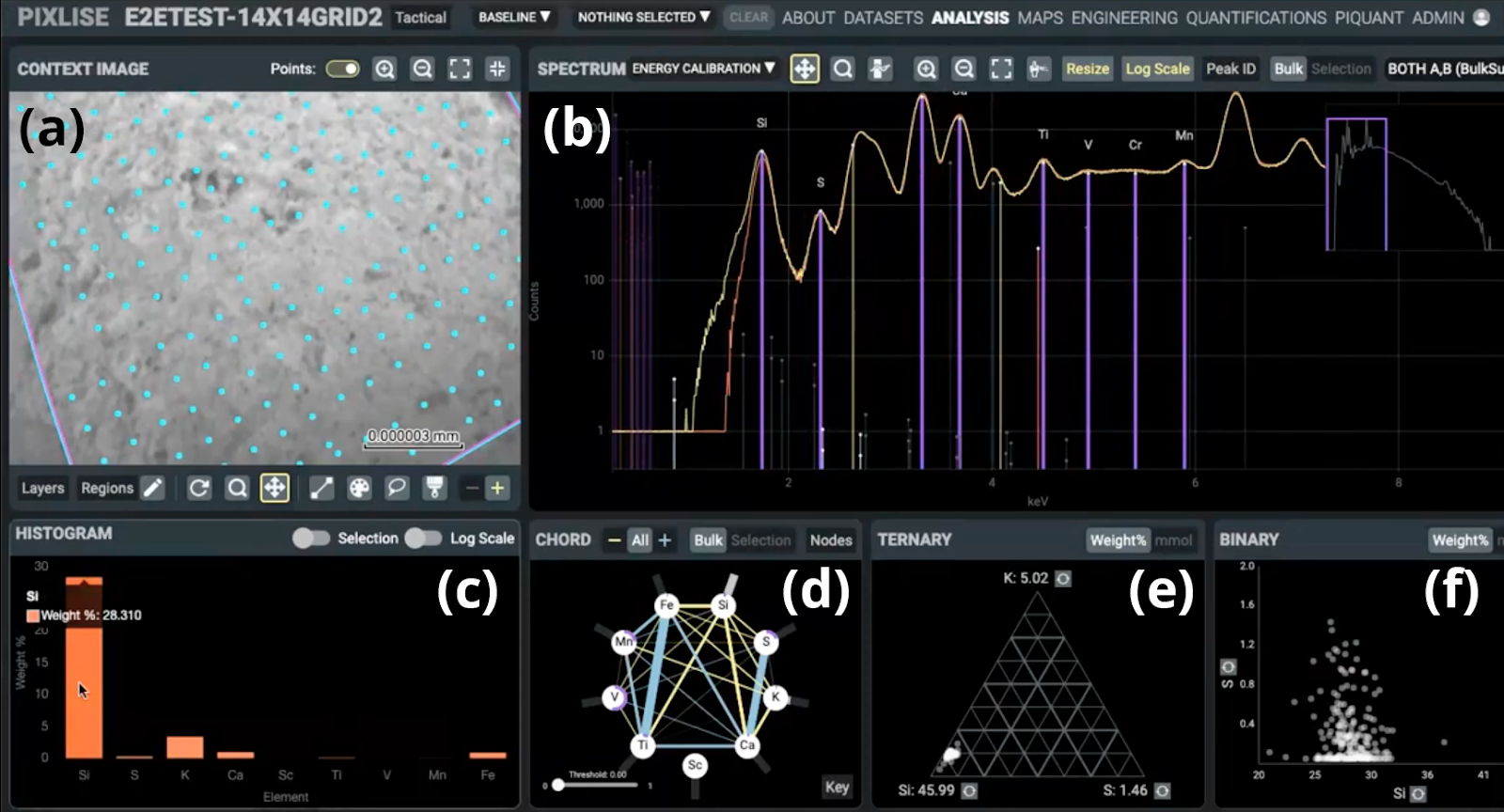}
    \caption{PIXLISE Interface}
    \label{fig:pixlise-interface}
\end{figure}

The current data visualization tool facilitates the data analysis workflow through multiple interaction techniques (see \autoref{fig:pixlise-interface} for the analysis tab). The main interaction touchpoint is the context image pane (\autoref{fig:pixlise-interface}a) that displays the discrete points in the microXRF data. Using this map display allows scientists to visually isolate, select, and analyze discrete geological features within an experiment site \cite{PIXELATE}. As a first step in their analysis, scientists evaluate the peaks in the returned spectra, and identify peaks that correspond to different elemental signatures. The spectrum pane (\autoref{fig:pixlise-interface}b) visualizes the energy levels as a function of wavelength, across each sample \cite{SpectralDensity}, and is examined individually or in bulk.

NASA has shown the instrument to be able to detect over 20 elements at 10 ppm \cite{PIXL4Scientists}. A spectroscopist drives the element identification process, computing the best fit for peaks using the Piquant algorithm. This process relies on a physics-based model to compute elemental weight percents from spectral peaks \cite{ELAM2002121}, and disambiguates them from diffraction, back-scatter and pile-up peaks, and background radiation.

Once the elements are identified, users can look at summary statistics using the histogram tool (\autoref{fig:pixlise-interface}c), which shows the quantified weight percentages of different elements in selected sample areas within the experiment. Scientists will then look to examine relationships between elemental abundance across the entire experiment area, and look to make associations of elemental abundance with the morphological structures in the rock. The chord diagram (\autoref{fig:pixlise-interface}d), the ternary diagram (\autoref{fig:pixlise-interface}e), and the binary diagram (\autoref{fig:pixlise-interface}f) support these investigations by showing the spatial covariance between quantified elements across the experiment. Stronger covariance suggests an underlying relationship that could point towards an underlying mineralogical cause.

While the PIXLISE tool supports the process of element assignment and quantification, our interviews with the stakeholders, (including the project lead UX Manager and a co-investigator research scientist), revealed that there are several pain points and needs in terms of selection, grouping, annotation, clustering, and comparison. First, our stakeholders shared that there is a need for supporting comparison of points or multiple clusters of samples, which is a crucial step in accurate mineral identification. Second, they shared that there is a need for doing multiple selections (individual or group) at the same time in order to compare the element distributions across different regions. Third, the scientists explained an underlying desire to have auto generated clusters with various algorithm options, so that they can semi-automatically create groups of the mapped area, instead of starting from scratch.Based on these needs and requirements, we developed PIXLISE-C. Our tool aims to address these gaps through extending the existing data analysis workflow.

\section{PIXLISE-C}

We present an interactive data visualization tool, PIXLISE-C, that extends the current system in place. We built our tool as a standalone application instead of integrated development, as the current architecture has several authentication layers that makes the development process difficult. However, our tool builds on the current application in terms of interface layout, interaction styles, and overall design to maintain familiarity.

\begin{figure}[h]
    \centering
    \includegraphics[width=\linewidth]{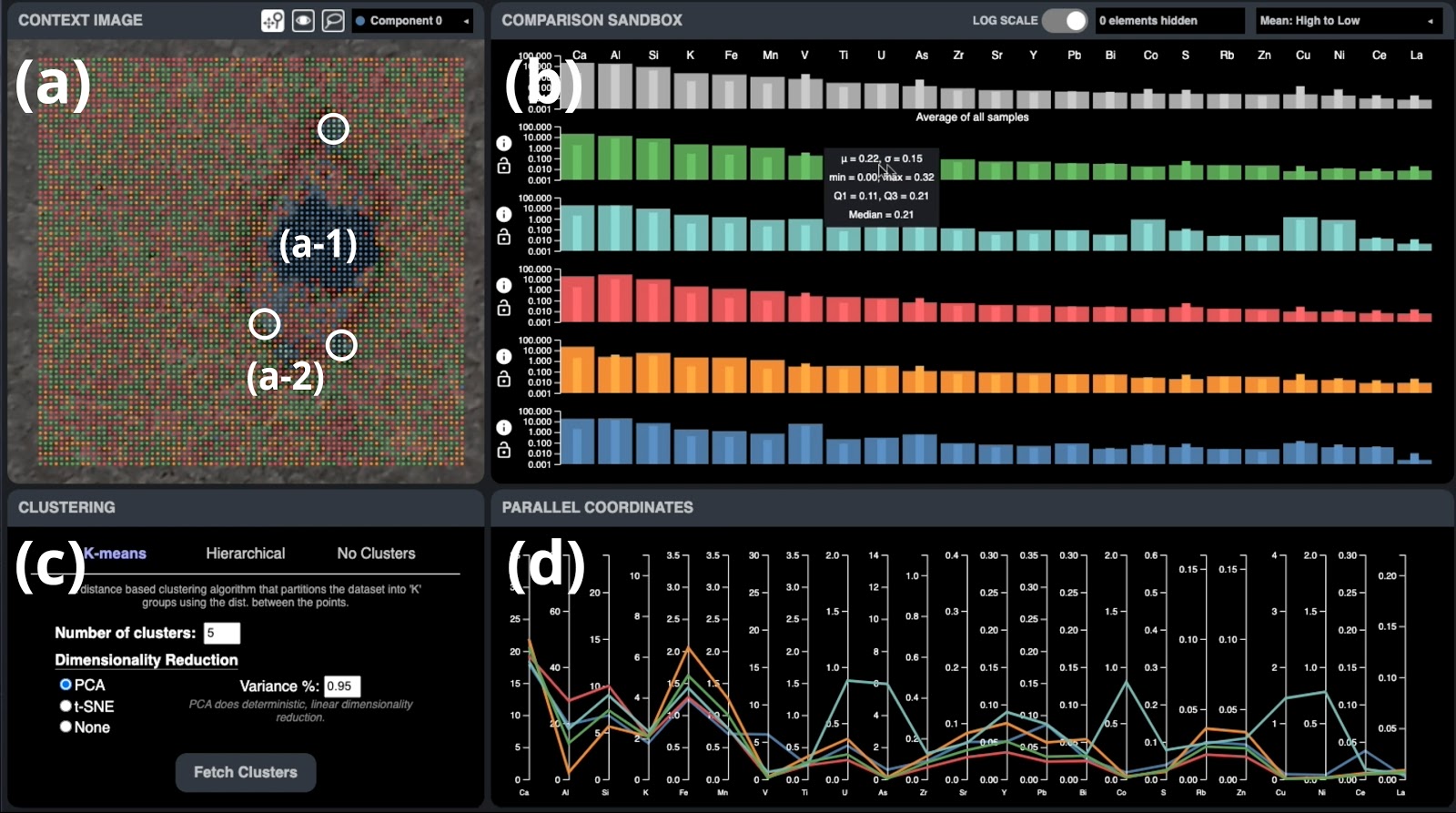}
    \caption{PIXLISE-C Interface}
    \label{fig:pixlise-c}
\end{figure}

PIXLISE-C interface consists of four main components: Context Image (\autoref{fig:pixlise-c}a), Comparison Sandbox (\autoref{fig:pixlise-c}b), Clustering (\autoref{fig:pixlise-c}c), and Parallel Coordinate Plots (\autoref{fig:pixlise-c}d). Below, we describe each component in detail.

\subsection{Context Image}
This 2D map is the main interaction touchpoint for the application. The visualization plots the collected data on the geographical context image as dot per point. In the demo interface, there are 6400 data points arranged into a grid based on their spatial coordinates: however, in different datasets it’s possible that the number and arrangement of points could differ. Overlaying the datapoint over the real terrain view allows scientists to select and analyze data in accordance with the context. There are three major actions the users can take in this view. First, using the navigation icon, they can pan and zoom (in or out). Second, using the examining tool (eye icon), they can hover over individual points to view related data in isolation, which is displayed in the Comparison Sandbox. In the examining tool, they will also be able to view any user annotations for that group, if they exist. Finally, the scientists can select and group the data points using the lasso tool and component dropdown menu. Clicking on “add item” creates a component group, and future selected data points will be displayed in that group’s color. Consequent lasso actions will add to the selection, while holding down shift will remove points from the selection. Detailed information about the groups in terms of element concentration is displayed in the Comparison Sandbox, so that the scientists can iterate on grouping data into clusters based on similarity.

\subsection{Comparison Sandbox}
\begin{figure}[h]
    \centering
    \begin{minipage}[b]{0.45\linewidth}
        \centering
        \includegraphics[width=\linewidth]{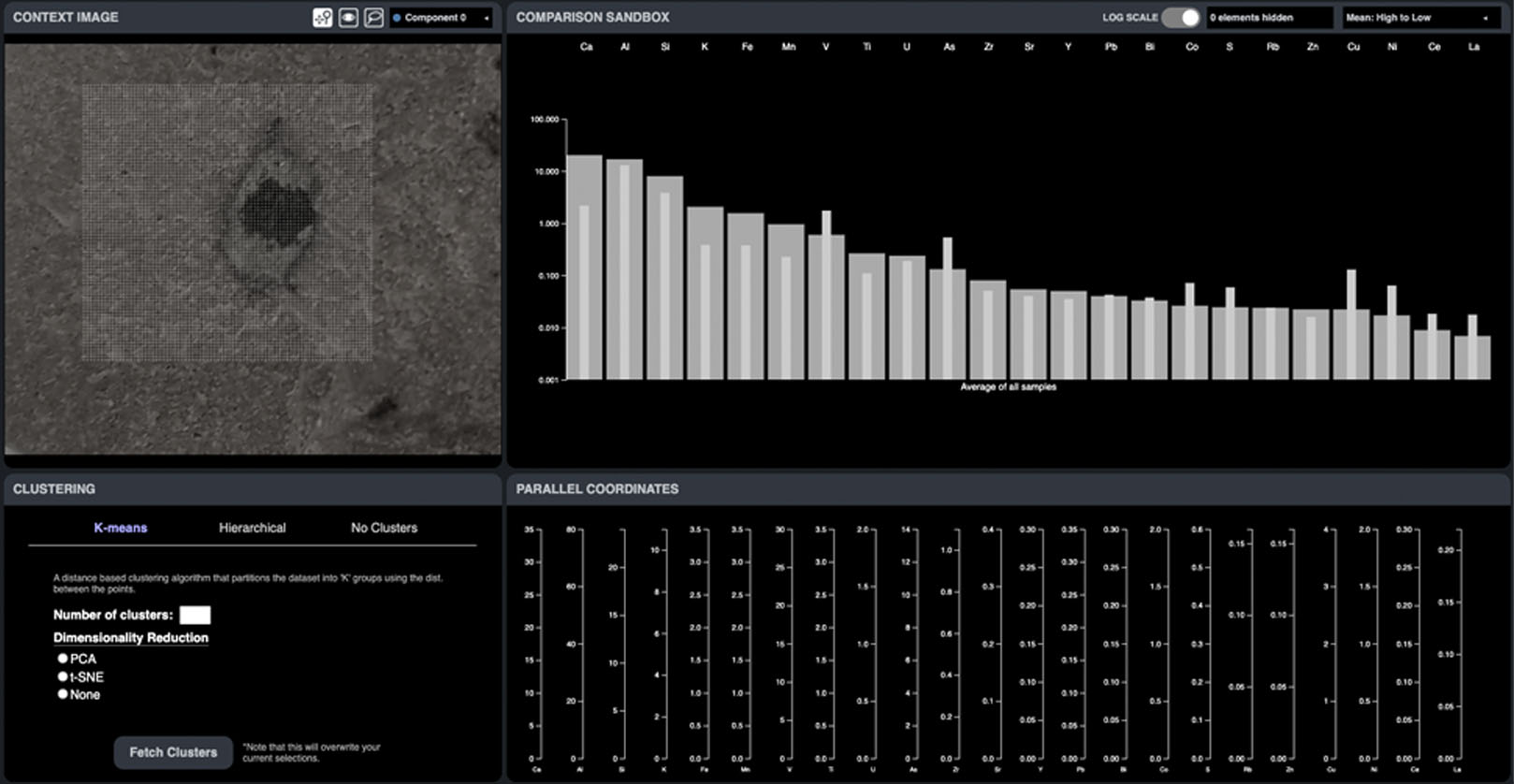}\\
        \subcaption{In the starting state, the sandbox contains a summary of all points}
    \end{minipage}\hfill%
    \begin{minipage}[b]{0.45\linewidth}
        \centering
        \includegraphics[width=\linewidth]{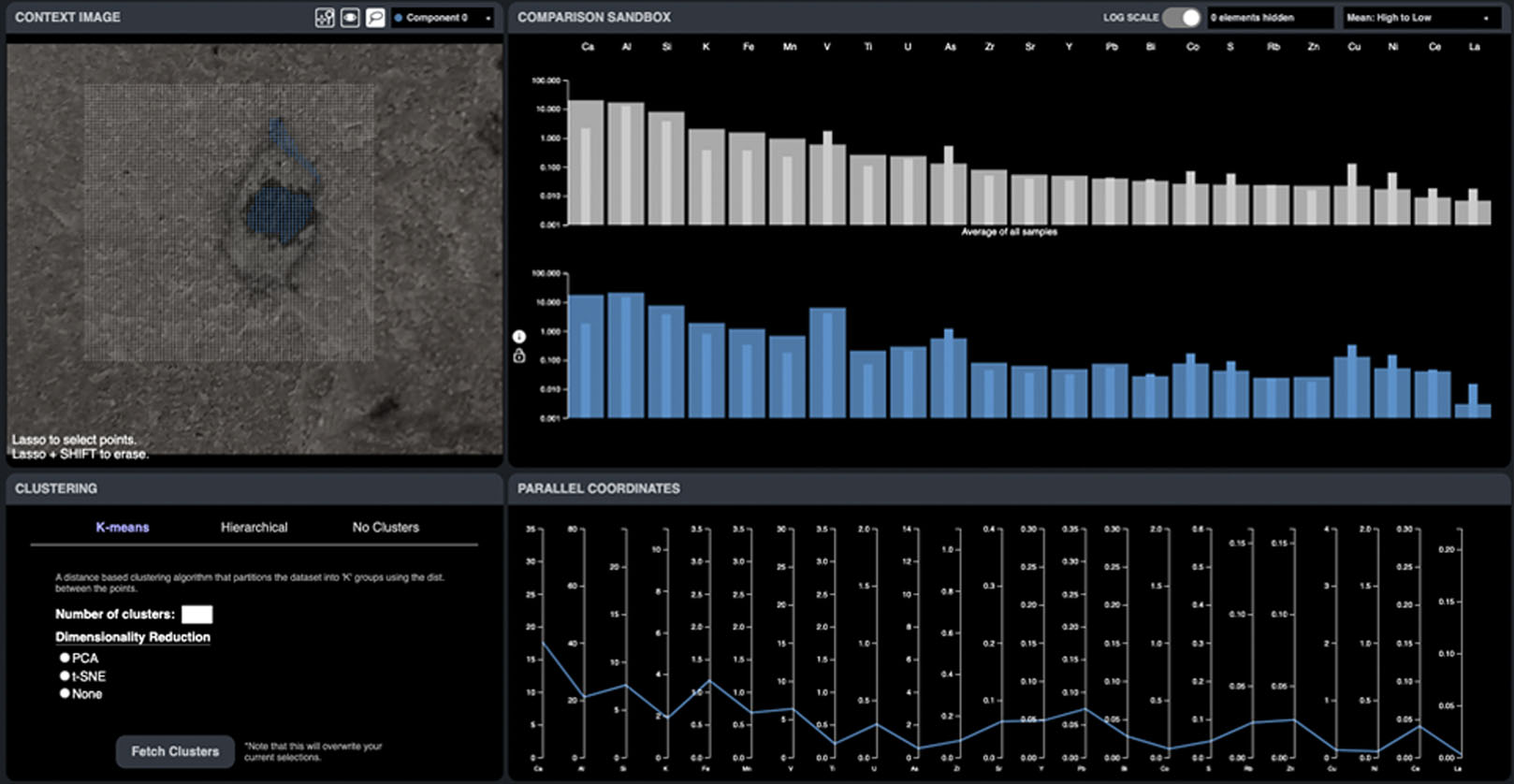}\\
        \subcaption{After selecting a group of blue points, the sandbox adds a histogram.}
    \end{minipage}
    \linebreak\linebreak
    \begin{minipage}[b]{0.45\linewidth}
        \centering
        \includegraphics[width=\linewidth]{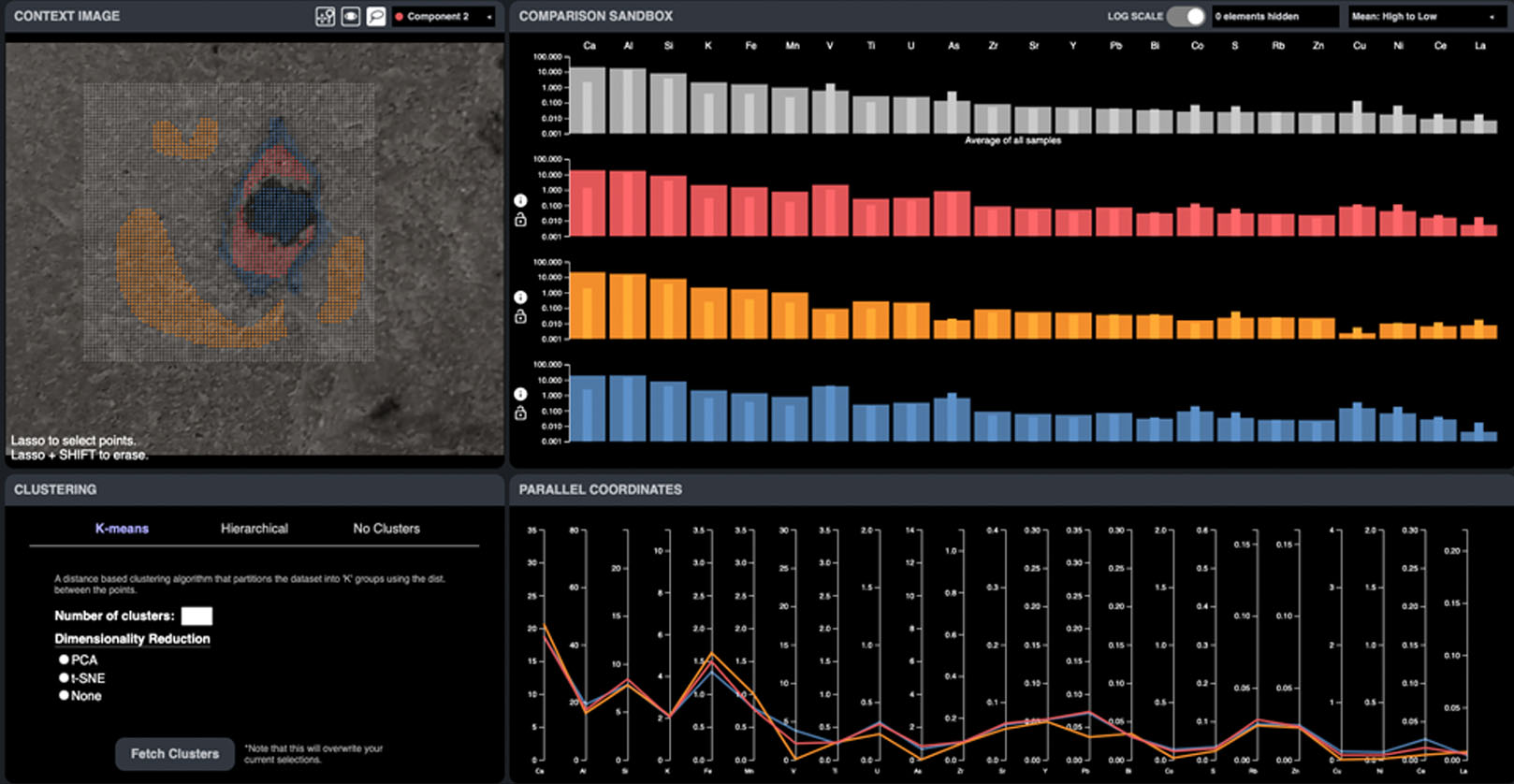}\\
        \subcaption{After selecting two more groups, the sandbox now has three more histograms.}
    \end{minipage}\hfill%
    \begin{minipage}[b]{0.45\linewidth}
        \centering
        \includegraphics[width=\linewidth]{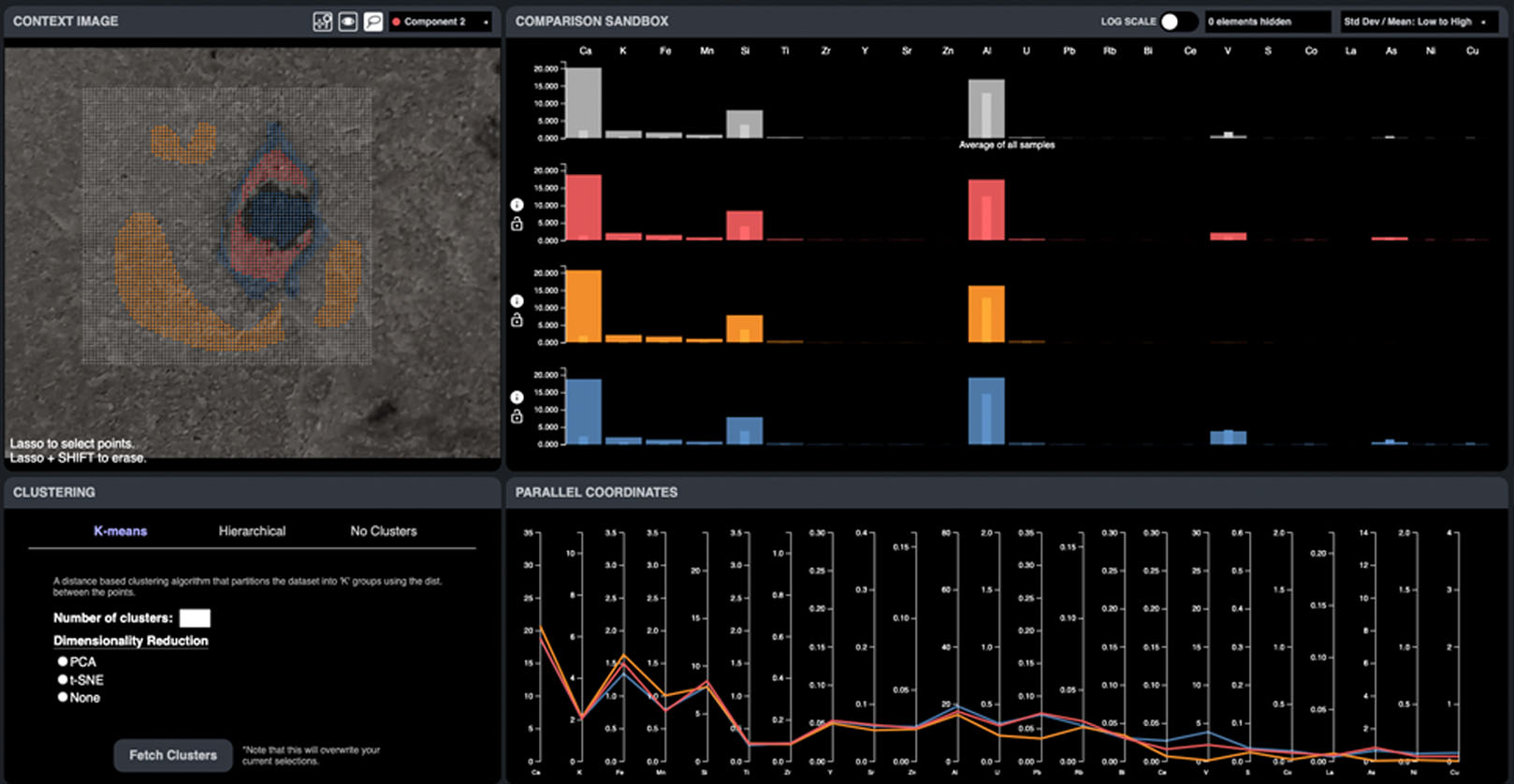}\\
        \subcaption{Histogram data can be transformed using filtering and sorting tools.}
    \end{minipage}%
    \label{fig:flipbook}
    \caption{Showing one possible workflow for PIXLISE-C}
\end{figure}
This view displays the element percentages in the selected data points as histograms. Instead of displaying one histogram at a time as in the previous implementation, our tool can display up to 20 histograms for the comparison of potential mineral compositions across clusters. When initiated the Comparison Sandbox displays the average element weights of all samples in the dataset on a logarithmic or linear scale. The element symbols are displayed as columns at the top and users can hide less important elements to focus on the most relevant ones. The histogram can be sorted based on the element abundance (using means, high to low or low to high) or the coefficient of variation (standard deviation divided by the mean). The histogram tooltip displays more detailed statistical information per element, including mean, standard deviation, min, max, Q1, Q3, and the median. This information is useful to help the user determine how the distribution of data looks. Min, max, Q1, Q3 and median are especially important if the distribution is not a normal distribution. Additionally, variance in an element is shown as an inner bar within each bar of the histogram, which informs whether selected data points demonstrate similar characteristics. We discussed showing variance on the bars with a line centered around the mean, but due to the large dataset and the presence of the log toggle, we believe our approach is able to convey the same information while being cleaner visually, especially in log scale.

\begin{figure}[h]
    \centering
    \includegraphics[width=\linewidth]{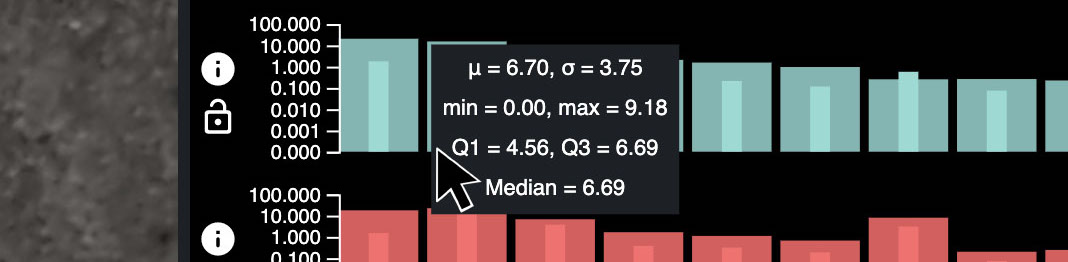}
    \caption{An example of a tooltip.}
    \label{fig:tooltip}
\end{figure}

As users create multiple groups using the selection and component creation tools in the Context Image view, the histograms in the Comparison Sandbox update to display multiple groups at once. Using histogram controls, users can lock and unlock a group before finalizing their selection. Additionally, they can annotate the group with insights. Once a group is annotated, hovering over the group in the Context Image view shows the annotations overlaid on top. Finally, when users want to examine how a single data point fits in a group, hovering over the data point in the Context Image view displays the element percentages of the individual data point across all histograms as overlaid ticks on bars.

\begin{figure}[h]
    \centering
    \includegraphics[width=\linewidth]{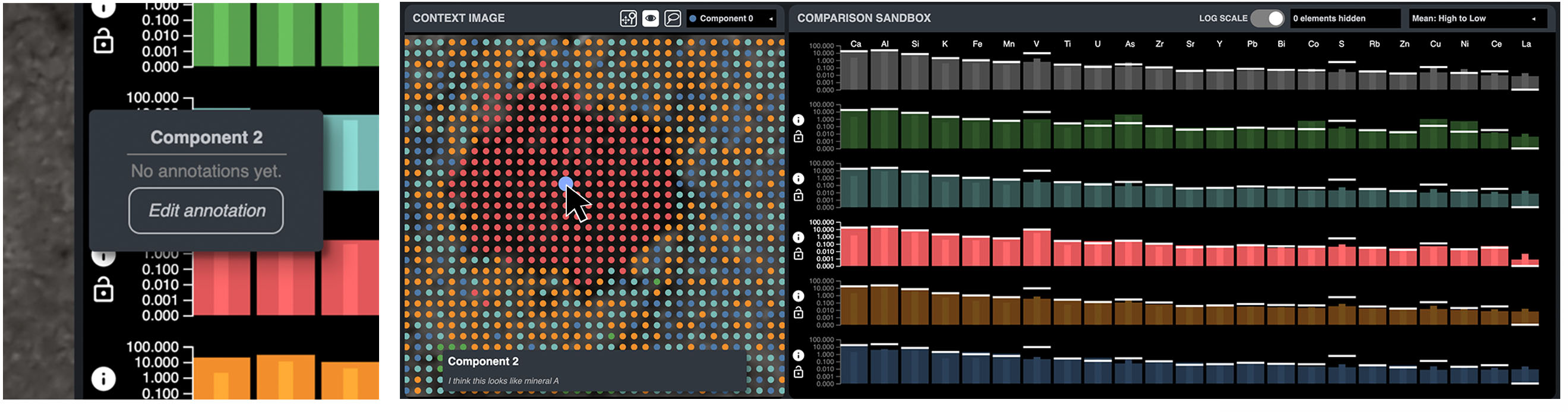}
    \caption{Left: How annotations are assigned. Right: when the examining tool view is selected, the context image displays the assigned annotation and the sandbox shows the stats of the point selected overlaid on all the groups.}
    \label{fig:compare}
\end{figure}

\subsection{Parallel Coordinates Plots}


Histograms are useful in characterizing groups through element abundance, but it can be difficult to distinguish fine grain differences among the data. To enable a more granular exploration, Parallel Coordinates Plots (PCP), as seen in \autoref{fig:pixlise-c}d, displays percent values based on the minimum and maximum range. Therefore, nuanced differences between groups can be seen per each element.

\subsection{Clustering}

\begin{figure}[h]
    \centering
   \begin{minipage}[b]{0.2\linewidth}
   \centering
        \includegraphics[width=\linewidth]{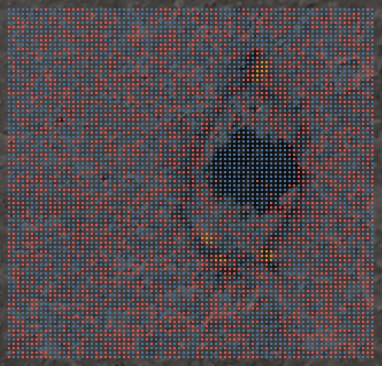}
        \subcaption{\textbf{Hierarchical}}
    \end{minipage}\hfill%
    \begin{minipage}[b]{0.2\linewidth}
    \centering
        \includegraphics[width=\linewidth]{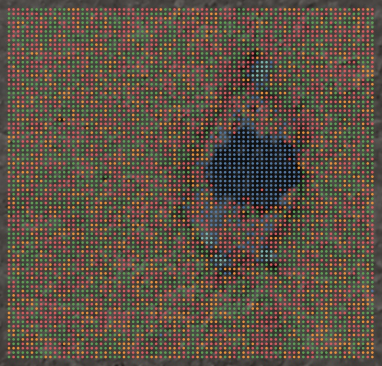}\\
        \subcaption{\textbf{K-means} \\ \emph{With PCA}}
    \end{minipage}\hfill%
    \begin{minipage}[b]{0.2\linewidth}
    \centering
        \includegraphics[width=\linewidth]{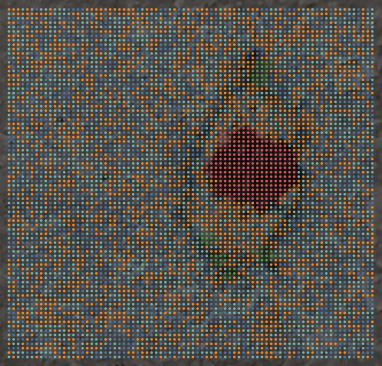}\\
        \subcaption{\textbf{K-means} \\ \emph{With t-SNE}}
    \end{minipage}\hfill%
    \begin{minipage}[b]{0.2\linewidth}
    \centering
        \includegraphics[width=\linewidth]{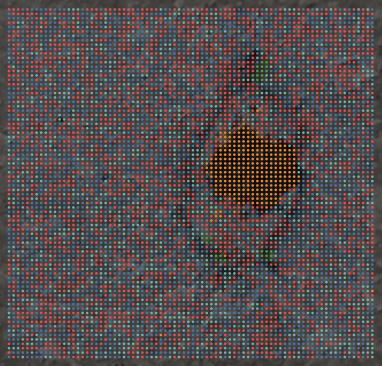}\\
        \subcaption{\textbf{K-means} \\ \emph{No reduction}}
    \end{minipage}%
    \caption{The results of different methods of clustering and dimensionality reduction, all set to generate 5 clusters. Notably, all of them are able to identify the crater in the middle but have subtle variations in how they group the points.}
    \label{fig:clusters}
\end{figure}

Scientists typically carry out the grouping of data points manually as they have extensive expertise in identifying geologic features in images. Our tool uses various ML techniques for auto generating clusters as starting points for a semi-automatic data analysis workflow. The clustering tool provides three different algorithms: k-means clustering, hierarchical clustering, and minimum maximum. For K-means, scientists can choose to apply dimensionality reduction using PCA or t-SNE. The clustering tool aims to enable scientists to try different clustering options and fine tune parameters for  finding patterns within the data and for creating initial groupings which they can manually edit afterwards.

\section{Technical Implementation}

We developed our application using React and deployed it using GitHub pages. Our code is open source \cite{Repo}. While React was used to develop the UI and coordinate the different components across the application, D3.js \cite{2011-d3}, was used to create the visualizations. 

For the clustering implementation, we use a Python backend to dynamically generate cluster information based on the many different hyperparameters that our app allows.The client app and the Python backend communicate using Flask. The backend itself is hosted in the cloud. Sklearn is used to perform all of our machine learning algorithms, including hierarchical clustering, t-SNE, PCA, and k-means. Our application also enables hyperparameters such as linkage, variance percentage, perplexity, and number of clusters, to be passed to the backend as well.

\section{Results}

Our tool imports a sample dataset called “King’s Court,” which has samples from a terrain with a crater-like geological feature. Using the clustering tool, we were able to test various algorithms and hyperparameters for grouping the dataset into meaningful subsets. Our explorative results show that both PCA and t-SNE were able to identify the large cluster around the crater (\autoref{fig:pixlise-c}a-1) with multiple discrete groups around its perimeters (\autoref{fig:pixlise-c}a-2). Once the initial groups were identified using the clustering algorithm, we used the hovering feature in the Context Image tools for examining individual data points within clusters. Based on the histogram information in the Comparison Sandbox, we were able to add or remove specific data points using the lasso tool. These groupings allow an efficient starting place for deciding what regions to compare and explore within the dataset. Thus the researcher can use PIXLISE-C to not only compare groupings they had already decided, but can also work with the program to discover new possible groupings that provide insight when compared.

\section{Discussion and Future Work}
Our work identifies information needs and requirements for mineral analysis tools. We operationalize the identified requirements in our proposed tool and present interaction techniques for the comparative analysis and clustering of microXRF spectroscopy data in the context of the PIXL project. Our tool PIXLISE-C enabled a semi-automatic data analysis workflow that enabled an augmented grouping of microXRF data based on element abundance. Using built-in statistical analyses, scientists can compare several clusters at once to fine tune and validate the grouping of the data. Initial feedback from our stakeholders suggested that tools that support selection and comparison are likely to improve data analysis workflows in the context of mineral exploration. Additional design requirements include documentation and annotation support for multi-user collaborative workflows.

Our work reveals several directions for future research. For the overall data analysis workflow, blending human and machine intelligence raises major research questions. How can we combine manually created groups with machine generated clusters to provide support scientists’ agency while removing tedious work? What kind of interaction techniques (e.g. overlays) might be suited for representing human and machine generated groups? Should the system suggest potential clusters to validate human insights through mixed initiative interaction? These are some questions we would like to explore going forward. In addition, sharing annotated datasets present an interesting direction for future research in terms of collaborative and asynchronous workflows. Potential techniques include exporting screenshots, sharing workspaces, and sharing the dataset to be opened with a different program or tool. The stakeholders expressed a need for a way to both present their findings in a visual manner but also in a text-based format that could plug into other scientific software. We will also explore the database requirements to allow saving and sharing.

Finally, our short term goals include the integration of PIXLISE-C into the current system used by NASA JPL. We will continue working with our collaborators to collect feedback on our tool through user testing, iterative development, and deployment.

\begin{acks}
We wish to thank the PIXL science team for supporting this research. The research was carried out, in part, by the Jet Propulsion Laboratory, California Institute of Technology, under a contract with the National Aeronautics and Space Administration (80NM0018D0004)
\end{acks}

\bibliographystyle{ACM-Reference-Format}
\bibliography{sample-base}


\begin{thebibliography}{22}


\ifx \showCODEN    \undefined \def \showCODEN     #1{\unskip}     \fi
\ifx \showDOI      \undefined \def \showDOI       #1{#1}\fi
\ifx \showISBNx    \undefined \def \showISBNx     #1{\unskip}     \fi
\ifx \showISBNxiii \undefined \def \showISBNxiii  #1{\unskip}     \fi
\ifx \showISSN     \undefined \def \showISSN      #1{\unskip}     \fi
\ifx \showLCCN     \undefined \def \showLCCN      #1{\unskip}     \fi
\ifx \shownote     \undefined \def \shownote      #1{#1}          \fi
\ifx \showarticletitle \undefined \def \showarticletitle #1{#1}   \fi
\ifx \showURL      \undefined \def \showURL       {\relax}        \fi
\providecommand\bibfield[2]{#2}
\providecommand\bibinfo[2]{#2}
\providecommand\natexlab[1]{#1}
\providecommand\showeprint[2][]{arXiv:#2}

\bibitem[\protect\citeauthoryear{Allwood, Grotzinger, Knoll, Burch, Anderson,
  Coleman, and Kanik}{Allwood et~al\mbox{.}}{2009}]%
        {Allwood9548}
\bibfield{author}{\bibinfo{person}{Abigail~C. Allwood},
  \bibinfo{person}{John~P. Grotzinger}, \bibinfo{person}{Andrew~H. Knoll},
  \bibinfo{person}{Ian~W. Burch}, \bibinfo{person}{Mark~S. Anderson},
  \bibinfo{person}{Max~L. Coleman}, {and} \bibinfo{person}{Isik Kanik}.}
  \bibinfo{year}{2009}\natexlab{}.
\newblock \showarticletitle{Controls on development and diversity of Early
  Archean stromatolites}.
\newblock \bibinfo{journal}{\emph{Proceedings of the National Academy of
  Sciences}} \bibinfo{volume}{106}, \bibinfo{number}{24}
  (\bibinfo{year}{2009}), \bibinfo{pages}{9548--9555}.
\newblock
\showISSN{0027-8424}
\urldef\tempurl%
\url{https://doi.org/10.1073/pnas.0903323106}
\showDOI{\tempurl}
\showeprint{https://www.pnas.org/content/106/24/9548.full.pdf}


\bibitem[\protect\citeauthoryear{Allwood, Wade, Foote, Elam, Hurowitz, Battel,
  Dawson, Denise, Ek, Gilbert, King, Liebe, Parker, Pedersen, Randall, Sharrow,
  Sondheim, Allen, Arnett, Au, Basset, Benn, Bousman, Braun, Calvet, Clark,
  Cinquini, Conaby, Conley, Davidoff, Delaney, Denver, Diaz, Doran, Ervin,
  Evans, Flannery, Gao, Gross, Grotzinger, Hannah, Harris, Harris, He,
  Heirwegh, Hernandez, Hertzberg, Hodyss, Holden, Hummel, Jadusingh,
  J{\o}rgensen, Kawamura, Kitiyakara, Kozaczek, Lambert, Lawson, Liu, Luchik,
  Macneal, Madsen, McLennan, McNally, Meras, Muller, Napoli, Naylor, Nemere,
  Ponomarev, Perez, Pootrakul, Romero, Rosas, Sachs, Schaefer, Schein,
  Setterfield, Singh, Song, Soria, Stek, Tallarida, Thompson, Tice, Timmermann,
  Torossian, Treiman, Tsai, Uckert, Villalvazo, Wang, Wilson, Worel, Zamani,
  Zappe, Zhong, and Zimmerman}{Allwood et~al\mbox{.}}{2020}]%
        {Allwood2020}
\bibfield{author}{\bibinfo{person}{Abigail~C. Allwood},
  \bibinfo{person}{Lawrence~A. Wade}, \bibinfo{person}{Marc~C. Foote},
  \bibinfo{person}{William~Timothy Elam}, \bibinfo{person}{Joel~A. Hurowitz},
  \bibinfo{person}{Steven Battel}, \bibinfo{person}{Douglas~E. Dawson},
  \bibinfo{person}{Robert~W. Denise}, \bibinfo{person}{Eric~M. Ek},
  \bibinfo{person}{Martin~S. Gilbert}, \bibinfo{person}{Matthew~E. King},
  \bibinfo{person}{Carl~Christian Liebe}, \bibinfo{person}{Todd Parker},
  \bibinfo{person}{David A.~K. Pedersen}, \bibinfo{person}{David~P. Randall},
  \bibinfo{person}{Robert~F. Sharrow}, \bibinfo{person}{Michael~E. Sondheim},
  \bibinfo{person}{George Allen}, \bibinfo{person}{Kenneth Arnett},
  \bibinfo{person}{Mitchell~H. Au}, \bibinfo{person}{Christophe Basset},
  \bibinfo{person}{Mathias Benn}, \bibinfo{person}{John~C. Bousman},
  \bibinfo{person}{David Braun}, \bibinfo{person}{Robert~J. Calvet},
  \bibinfo{person}{Benton Clark}, \bibinfo{person}{Luca Cinquini},
  \bibinfo{person}{Sterling Conaby}, \bibinfo{person}{Henry~A. Conley},
  \bibinfo{person}{Scott Davidoff}, \bibinfo{person}{Jenna Delaney},
  \bibinfo{person}{Troelz Denver}, \bibinfo{person}{Ernesto Diaz},
  \bibinfo{person}{Gary~B. Doran}, \bibinfo{person}{Joan Ervin},
  \bibinfo{person}{Michael Evans}, \bibinfo{person}{David~O. Flannery},
  \bibinfo{person}{Ning Gao}, \bibinfo{person}{Johannes Gross},
  \bibinfo{person}{John Grotzinger}, \bibinfo{person}{Brett Hannah},
  \bibinfo{person}{Jackson~T. Harris}, \bibinfo{person}{Cathleen~M. Harris},
  \bibinfo{person}{Yejun He}, \bibinfo{person}{Christopher~M. Heirwegh},
  \bibinfo{person}{Christina Hernandez}, \bibinfo{person}{Eric Hertzberg},
  \bibinfo{person}{Robert~P. Hodyss}, \bibinfo{person}{James~R. Holden},
  \bibinfo{person}{Christopher Hummel}, \bibinfo{person}{Matthew~A. Jadusingh},
  \bibinfo{person}{John~L. J{\o}rgensen}, \bibinfo{person}{Jonathan~H.
  Kawamura}, \bibinfo{person}{Amarit Kitiyakara}, \bibinfo{person}{Kris
  Kozaczek}, \bibinfo{person}{James~L. Lambert}, \bibinfo{person}{Peter~R.
  Lawson}, \bibinfo{person}{Yang Liu}, \bibinfo{person}{Thomas~S. Luchik},
  \bibinfo{person}{Kristen~M. Macneal}, \bibinfo{person}{Soren~N. Madsen},
  \bibinfo{person}{Scott~M. McLennan}, \bibinfo{person}{Patrick McNally},
  \bibinfo{person}{Patrick~L. Meras}, \bibinfo{person}{Richard~E. Muller},
  \bibinfo{person}{Jamie Napoli}, \bibinfo{person}{Bret~J. Naylor},
  \bibinfo{person}{Peter Nemere}, \bibinfo{person}{Igor Ponomarev},
  \bibinfo{person}{Raul~M. Perez}, \bibinfo{person}{Napat Pootrakul},
  \bibinfo{person}{Raul~A. Romero}, \bibinfo{person}{Rogelio Rosas},
  \bibinfo{person}{Jared Sachs}, \bibinfo{person}{Rembrandt~T. Schaefer},
  \bibinfo{person}{Michael~E. Schein}, \bibinfo{person}{Timothy~P.
  Setterfield}, \bibinfo{person}{Vritika Singh}, \bibinfo{person}{Eugenie
  Song}, \bibinfo{person}{Mary~M. Soria}, \bibinfo{person}{Paul~C. Stek},
  \bibinfo{person}{Nicholas~R. Tallarida}, \bibinfo{person}{David~R. Thompson},
  \bibinfo{person}{Michael~M. Tice}, \bibinfo{person}{Lars Timmermann},
  \bibinfo{person}{Violet Torossian}, \bibinfo{person}{Allan Treiman},
  \bibinfo{person}{Shihchuan Tsai}, \bibinfo{person}{Kyle Uckert},
  \bibinfo{person}{Juan Villalvazo}, \bibinfo{person}{Mandy Wang},
  \bibinfo{person}{Daniel~W. Wilson}, \bibinfo{person}{Shana~C. Worel},
  \bibinfo{person}{Payam Zamani}, \bibinfo{person}{Mike Zappe},
  \bibinfo{person}{Fang Zhong}, {and} \bibinfo{person}{Richard Zimmerman}.}
  \bibinfo{year}{2020}\natexlab{}.
\newblock \showarticletitle{PIXL: Planetary Instrument for X-Ray
  Lithochemistry}.
\newblock \bibinfo{journal}{\emph{Space Science Reviews}}
  \bibinfo{volume}{216}, \bibinfo{number}{8} (\bibinfo{date}{19 Nov}
  \bibinfo{year}{2020}), \bibinfo{pages}{134}.
\newblock
\showISSN{1572-9672}
\urldef\tempurl%
\url{https://doi.org/10.1007/s11214-020-00767-7}
\showDOI{\tempurl}


\bibitem[\protect\citeauthoryear{Banfield, Moreau, Chan, Welch, and
  Little}{Banfield et~al\mbox{.}}{2001}]%
        {Banfield2001}
\bibfield{author}{\bibinfo{person}{Jillian~F. Banfield},
  \bibinfo{person}{John~W. Moreau}, \bibinfo{person}{Clara~S. Chan},
  \bibinfo{person}{Susan~A. Welch}, {and} \bibinfo{person}{Brenda Little}.}
  \bibinfo{year}{2001}\natexlab{}.
\newblock \showarticletitle{Mineralogical Biosignatures and the Search for Life
  on Mars}.
\newblock \bibinfo{journal}{\emph{Astrobiology}} \bibinfo{volume}{1},
  \bibinfo{number}{4} (\bibinfo{year}{2001}), \bibinfo{pages}{447--465}.
\newblock
\urldef\tempurl%
\url{https://doi.org/10.1089/153110701753593856}
\showDOI{\tempurl}
\showeprint{https://doi.org/10.1089/153110701753593856}
\newblock
\shownote{PMID: 12448978.}


\bibitem[\protect\citeauthoryear{Bostock, Ogievetsky, and Heer}{Bostock
  et~al\mbox{.}}{2011}]%
        {2011-d3}
\bibfield{author}{\bibinfo{person}{Michael Bostock}, \bibinfo{person}{Vadim
  Ogievetsky}, {and} \bibinfo{person}{Jeffrey Heer}.}
  \bibinfo{year}{2011}\natexlab{}.
\newblock \showarticletitle{D3: Data-Driven Documents}.
\newblock \bibinfo{journal}{\emph{IEEE Trans. Visualization \& Comp. Graphics
  (Proc. InfoVis)}} (\bibinfo{year}{2011}).
\newblock
\urldef\tempurl%
\url{http://idl.cs.washington.edu/papers/d3}
\showURL{%
\tempurl}


\bibitem[\protect\citeauthoryear{Elam, Ravel, and Sieber}{Elam
  et~al\mbox{.}}{2002}]%
        {ELAM2002121}
\bibfield{author}{\bibinfo{person}{W.T. Elam}, \bibinfo{person}{B.D. Ravel},
  {and} \bibinfo{person}{J.R. Sieber}.} \bibinfo{year}{2002}\natexlab{}.
\newblock \showarticletitle{A new atomic database for X-ray spectroscopic
  calculations}.
\newblock \bibinfo{journal}{\emph{Radiation Physics and Chemistry}}
  \bibinfo{volume}{63}, \bibinfo{number}{2} (\bibinfo{year}{2002}),
  \bibinfo{pages}{121--128}.
\newblock
\showISSN{0969-806X}
\urldef\tempurl%
\url{https://doi.org/10.1016/S0969-806X(01)00227-4}
\showDOI{\tempurl}


\bibitem[\protect\citeauthoryear{Farley, Williford, Stack, Bhartia, Chen,
  de la Torre, Hand, Goreva, Herd, Hueso, Liu, Maki, Martinez, Moeller,
  Nelessen, Newman, Nunes, Ponce, Spanovich, Willis, Beegle, Bell, Brown,
  Hamran, Hurowitz, Maurice, Paige, Rodriguez-Manfredi, Schulte, and
  Wiens}{Farley et~al\mbox{.}}{2020}]%
        {Farley2020}
\bibfield{author}{\bibinfo{person}{Kenneth~A. Farley},
  \bibinfo{person}{Kenneth~H. Williford}, \bibinfo{person}{Kathryn~M. Stack},
  \bibinfo{person}{Rohit Bhartia}, \bibinfo{person}{Al Chen},
  \bibinfo{person}{Manuel de la Torre}, \bibinfo{person}{Kevin Hand},
  \bibinfo{person}{Yulia Goreva}, \bibinfo{person}{Christopher D.~K. Herd},
  \bibinfo{person}{Ricardo Hueso}, \bibinfo{person}{Yang Liu},
  \bibinfo{person}{Justin~N. Maki}, \bibinfo{person}{German Martinez},
  \bibinfo{person}{Robert~C. Moeller}, \bibinfo{person}{Adam Nelessen},
  \bibinfo{person}{Claire~E. Newman}, \bibinfo{person}{Daniel Nunes},
  \bibinfo{person}{Adrian Ponce}, \bibinfo{person}{Nicole Spanovich},
  \bibinfo{person}{Peter~A. Willis}, \bibinfo{person}{Luther~W. Beegle},
  \bibinfo{person}{James~F. Bell}, \bibinfo{person}{Adrian~J. Brown},
  \bibinfo{person}{Svein-Erik Hamran}, \bibinfo{person}{Joel~A. Hurowitz},
  \bibinfo{person}{Sylvestre Maurice}, \bibinfo{person}{David~A. Paige},
  \bibinfo{person}{Jose~A. Rodriguez-Manfredi}, \bibinfo{person}{Mitch
  Schulte}, {and} \bibinfo{person}{Roger~C. Wiens}.}
  \bibinfo{year}{2020}\natexlab{}.
\newblock \showarticletitle{Mars 2020 Mission Overview}.
\newblock \bibinfo{journal}{\emph{Space Science Reviews}}
  \bibinfo{volume}{216}, \bibinfo{number}{8} (\bibinfo{date}{03 Dec}
  \bibinfo{year}{2020}), \bibinfo{pages}{142}.
\newblock
\showISSN{1572-9672}
\urldef\tempurl%
\url{https://doi.org/10.1007/s11214-020-00762-y}
\showDOI{\tempurl}


\bibitem[\protect\citeauthoryear{Flannery, Davidoff, Tice, Allwood, Elam,
  Heirwegh, Hurowitz, Liu, and Nemere}{Flannery et~al\mbox{.}}{[n.d.]}]%
        {PIXLISECOSPAR2021}
\bibfield{author}{\bibinfo{person}{David Flannery}, \bibinfo{person}{Scott
  Davidoff}, \bibinfo{person}{Michael~M. Tice}, \bibinfo{person}{Abigail~C.
  Allwood}, \bibinfo{person}{William~Timothy Elam},
  \bibinfo{person}{Christopher~M. Heirwegh}, \bibinfo{person}{Joel~A.
  Hurowitz}, \bibinfo{person}{Yang Liu}, {and} \bibinfo{person}{Peter Nemere}.}
  \bibinfo{year}{[n.d.]}\natexlab{}.
\newblock \showarticletitle{Increasing Efficiency of Mars 2020 Rover Operations
  via Novel Data Analysis Software for the Planetary Instrument for X-ray
  Lithochemistry (PIXL)}.
\newblock \bibinfo{journal}{\emph{Proceedings of the 2021 Committee on Space
  Research (COSPAR) Scientific Assembly}}  \bibinfo{volume}{43}
  (\bibinfo{year}{[n.\,d.]}).
\newblock
Issue B0.2.
\urldef\tempurl%
\url{https://www.cospar-assembly.org/user/download.php?id=28088&type=abstract&section=congressbrowser}
\showURL{%
\tempurl}


\bibitem[\protect\citeauthoryear{Horvitz}{Horvitz}{1999}]%
        {Horvitz}
\bibfield{author}{\bibinfo{person}{Eric Horvitz}.}
  \bibinfo{year}{1999}\natexlab{}.
\newblock \showarticletitle{Principles of Mixed-Initiative User Interfaces}. In
  \bibinfo{booktitle}{\emph{Proceedings of the SIGCHI Conference on Human
  Factors in Computing Systems}} (Pittsburgh, Pennsylvania, USA)
  \emph{(\bibinfo{series}{CHI '99})}. \bibinfo{publisher}{Association for
  Computing Machinery}, \bibinfo{address}{New York, NY, USA},
  \bibinfo{pages}{159–166}.
\newblock
\showISBNx{0201485591}
\urldef\tempurl%
\url{https://doi.org/10.1145/302979.303030}
\showDOI{\tempurl}


\bibitem[\protect\citeauthoryear{McLennan, Bell, Calvin, Christensen, Clark,
  {de Souza}, Farmer, Farrand, Fike, Gellert, Ghosh, Glotch, Grotzinger, Hahn,
  Herkenhoff, Hurowitz, Johnson, Johnson, Jolliff, Klingelhöfer, Knoll,
  Learner, Malin, McSween, Pocock, Ruff, Soderblom, Squyres, Tosca, Watters,
  Wyatt, and Yen}{McLennan et~al\mbox{.}}{2005}]%
        {MCLENNAN200595}
\bibfield{author}{\bibinfo{person}{S.M. McLennan}, \bibinfo{person}{J.F. Bell},
  \bibinfo{person}{W.M. Calvin}, \bibinfo{person}{P.R. Christensen},
  \bibinfo{person}{B.C. Clark}, \bibinfo{person}{P.A. {de Souza}},
  \bibinfo{person}{J. Farmer}, \bibinfo{person}{W.H. Farrand},
  \bibinfo{person}{D.A. Fike}, \bibinfo{person}{R. Gellert},
  \bibinfo{person}{A. Ghosh}, \bibinfo{person}{T.D. Glotch},
  \bibinfo{person}{J.P. Grotzinger}, \bibinfo{person}{B. Hahn},
  \bibinfo{person}{K.E. Herkenhoff}, \bibinfo{person}{J.A. Hurowitz},
  \bibinfo{person}{J.R. Johnson}, \bibinfo{person}{S.S. Johnson},
  \bibinfo{person}{B. Jolliff}, \bibinfo{person}{G. Klingelhöfer},
  \bibinfo{person}{A.H. Knoll}, \bibinfo{person}{Z. Learner},
  \bibinfo{person}{M.C. Malin}, \bibinfo{person}{H.Y. McSween},
  \bibinfo{person}{J. Pocock}, \bibinfo{person}{S.W. Ruff},
  \bibinfo{person}{L.A. Soderblom}, \bibinfo{person}{S.W. Squyres},
  \bibinfo{person}{N.J. Tosca}, \bibinfo{person}{W.A. Watters},
  \bibinfo{person}{M.B. Wyatt}, {and} \bibinfo{person}{A. Yen}.}
  \bibinfo{year}{2005}\natexlab{}.
\newblock \showarticletitle{Provenance and diagenesis of the evaporite-bearing
  Burns formation, Meridiani Planum, Mars}.
\newblock \bibinfo{journal}{\emph{Earth and Planetary Science Letters}}
  \bibinfo{volume}{240}, \bibinfo{number}{1} (\bibinfo{year}{2005}),
  \bibinfo{pages}{95--121}.
\newblock
\showISSN{0012-821X}
\urldef\tempurl%
\url{https://doi.org/10.1016/j.epsl.2005.09.041}
\showDOI{\tempurl}
\newblock
\shownote{Sedimentary Geology at Meridiani Planum, Mars.}


\bibitem[\protect\citeauthoryear{NASA}{NASA}{2020}]%
        {PIXL4Scientists}
\bibfield{author}{\bibinfo{person}{NASA}.} \bibinfo{year}{2020}\natexlab{}.
\newblock \bibinfo{title}{PIXL for Scientists}.
\newblock
\newblock
\urldef\tempurl%
\url{https://mars.nasa.gov/mars2020/spacecraft/instruments/pixl/for-scientists/}
\showURL{%
\tempurl}


\bibitem[\protect\citeauthoryear{NASA}{NASA}{2021}]%
        {PIXLISEwww}
\bibfield{author}{\bibinfo{person}{NASA}.} \bibinfo{year}{2021}\natexlab{}.
\newblock \bibinfo{title}{PIXLISE Application}.
\newblock
\newblock
\urldef\tempurl%
\url{https://www.pixlise.org/}
\showURL{%
\tempurl}


\bibitem[\protect\citeauthoryear{Rieder, Economou, W{\"a}nke, Turkevich, Crisp,
  Br{\"u}ckner, Dreibus, and McSween}{Rieder et~al\mbox{.}}{1997}]%
        {Rieder1771}
\bibfield{author}{\bibinfo{person}{R. Rieder}, \bibinfo{person}{T. Economou},
  \bibinfo{person}{H. W{\"a}nke}, \bibinfo{person}{A. Turkevich},
  \bibinfo{person}{J. Crisp}, \bibinfo{person}{J. Br{\"u}ckner},
  \bibinfo{person}{G. Dreibus}, {and} \bibinfo{person}{H.~Y. McSween}.}
  \bibinfo{year}{1997}\natexlab{}.
\newblock \showarticletitle{The Chemical Composition of Martian Soil and Rocks
  Returned by the Mobile Alpha Proton X-ray Spectrometer: Preliminary Results
  from the X-ray Mode}.
\newblock \bibinfo{journal}{\emph{Science}} \bibinfo{volume}{278},
  \bibinfo{number}{5344} (\bibinfo{year}{1997}), \bibinfo{pages}{1771--1774}.
\newblock
\showISSN{0036-8075}
\urldef\tempurl%
\url{https://doi.org/10.1126/science.278.5344.1771}
\showDOI{\tempurl}
\showeprint{https://science.sciencemag.org/content/278/5344/1771.1.full.pdf}


\bibitem[\protect\citeauthoryear{Ruff and Farmer}{Ruff and Farmer}{2016}]%
        {Ruff2016}
\bibfield{author}{\bibinfo{person}{Steven~W. Ruff} {and}
  \bibinfo{person}{Jack~D. Farmer}.} \bibinfo{year}{2016}\natexlab{}.
\newblock \showarticletitle{Silica deposits on Mars with features resembling
  hot spring biosignatures at El Tatio in Chile}.
\newblock \bibinfo{journal}{\emph{Nature Communications}} \bibinfo{volume}{7},
  \bibinfo{number}{1} (\bibinfo{date}{17 Nov} \bibinfo{year}{2016}),
  \bibinfo{pages}{13554}.
\newblock
\showISSN{2041-1723}
\urldef\tempurl%
\url{https://doi.org/10.1038/ncomms13554}
\showDOI{\tempurl}


\bibitem[\protect\citeauthoryear{Sanders and Stappers}{Sanders and
  Stappers}{2008}]%
        {sanders2008co}
\bibfield{author}{\bibinfo{person}{Elizabeth B-N Sanders} {and}
  \bibinfo{person}{Pieter~Jan Stappers}.} \bibinfo{year}{2008}\natexlab{}.
\newblock \showarticletitle{Co-creation and the new landscapes of design}.
\newblock \bibinfo{journal}{\emph{Co-design}} \bibinfo{volume}{4},
  \bibinfo{number}{1} (\bibinfo{year}{2008}), \bibinfo{pages}{5--18}.
\newblock
\urldef\tempurl%
\url{https://doi.org/10.1080/15710880701875068}
\showDOI{\tempurl}


\bibitem[\protect\citeauthoryear{Saunders, Arvidson, Badhwar, Boynton,
  Christensen, Cucinotta, Feldman, Gibbs, Kloss, Landano, Mase, McSmith, Meyer,
  Mitrofanov, Pace, Plaut, Sidney, Spencer, Thompson, and Zeitlin}{Saunders
  et~al\mbox{.}}{2004}]%
        {Saunders2001}
\bibfield{author}{\bibinfo{person}{R.S. Saunders}, \bibinfo{person}{Raymond
  Arvidson}, \bibinfo{person}{G.D. Badhwar}, \bibinfo{person}{William Boynton},
  \bibinfo{person}{P.R. Christensen}, \bibinfo{person}{Francis Cucinotta},
  \bibinfo{person}{W.C. Feldman}, \bibinfo{person}{R.G. Gibbs},
  \bibinfo{person}{C. Kloss}, \bibinfo{person}{M.R. Landano},
  \bibinfo{person}{R.A. Mase}, \bibinfo{person}{G.W. McSmith},
  \bibinfo{person}{Michael Meyer}, \bibinfo{person}{I.G. Mitrofanov},
  \bibinfo{person}{G.D. Pace}, \bibinfo{person}{J.J. Plaut},
  \bibinfo{person}{W.P. Sidney}, \bibinfo{person}{David Spencer},
  \bibinfo{person}{Thomas Thompson}, {and} \bibinfo{person}{Cary Zeitlin}.}
  \bibinfo{year}{2004}\natexlab{}.
\newblock \showarticletitle{2001 Mars Odyssey Mission Summary}.
\newblock \bibinfo{journal}{\emph{Space Science Reviews}}
  \bibinfo{volume}{110} (\bibinfo{date}{01} \bibinfo{year}{2004}),
  \bibinfo{pages}{1--36}.
\newblock
\showISBNx{978-94-015-6958-3}
\urldef\tempurl%
\url{https://doi.org/10.1023/B:SPAC.0000021006.84299.18}
\showDOI{\tempurl}


\bibitem[\protect\citeauthoryear{Schurman, Nair, Davidoff, Galvin, Allwood,
  Liu, Flannery, Hodyss, Lombeyda, Hendrie, Mushkin, and Heirwegh}{Schurman
  et~al\mbox{.}}{2019}]%
        {PIXELATE}
\bibfield{author}{\bibinfo{person}{David Schurman}, \bibinfo{person}{Pooja
  Nair}, \bibinfo{person}{Scott Davidoff}, \bibinfo{person}{Adrian Galvin},
  \bibinfo{person}{Abigail Allwood}, \bibinfo{person}{Yang Liu},
  \bibinfo{person}{David Flannery}, \bibinfo{person}{Robert~P. Hodyss},
  \bibinfo{person}{Santiago~V. Lombeyda}, \bibinfo{person}{Maggie Hendrie},
  \bibinfo{person}{Hillary Mushkin}, {and} \bibinfo{person}{Christopher~P.
  Heirwegh}.} \bibinfo{year}{2019}\natexlab{}.
\newblock \showarticletitle{PIXELATE: Novel visualization and computational
  methods for the analysis of astrobiological spectroscopy data}.
\newblock \bibinfo{journal}{\emph{Proceedings of the 2019 Astrobiology Science
  Conference (AbSciCon)}}.
\newblock
\urldef\tempurl%
\url{https://agu.confex.com/agu/abscicon19/prelim.cgi/Paper/482995}
\showURL{%
\tempurl}


\bibitem[\protect\citeauthoryear{{Shneiderman}}{{Shneiderman}}{1996}]%
        {Shneiderman}
\bibfield{author}{\bibinfo{person}{B. {Shneiderman}}.}
  \bibinfo{year}{1996}\natexlab{}.
\newblock \showarticletitle{The eyes have it: a task by data type taxonomy for
  information visualizations}. In \bibinfo{booktitle}{\emph{Proceedings 1996
  IEEE Symposium on Visual Languages}}. \bibinfo{pages}{336--343}.
\newblock
\urldef\tempurl%
\url{https://doi.org/10.1109/VL.1996.545307}
\showDOI{\tempurl}


\bibitem[\protect\citeauthoryear{to~Discovery~Program}{to~Discovery~Program}{2019}]%
        {PixlDatavis}
\bibfield{author}{\bibinfo{person}{Data to Discovery~Program}.}
  \bibinfo{year}{2019}\natexlab{}.
\newblock \bibinfo{title}{PIXLISE Mars Rock Sample Investigation}.
\newblock
\newblock
\urldef\tempurl%
\url{https://datavis.caltech.edu/projects/pixl/}
\showURL{%
\tempurl}


\bibitem[\protect\citeauthoryear{VanBommel, Gellert, Berger, Campbell,
  Thompson, Edgett, McBride, Minitti, Pradler, and Boyd}{VanBommel
  et~al\mbox{.}}{2016}]%
        {VanBommel2016}
\bibfield{author}{\bibinfo{person}{Scott~J. VanBommel}, \bibinfo{person}{Ralf
  Gellert}, \bibinfo{person}{Jeff~A. Berger}, \bibinfo{person}{John~L.
  Campbell}, \bibinfo{person}{Lucy~M. Thompson}, \bibinfo{person}{Kenneth~S.
  Edgett}, \bibinfo{person}{Marie~J. McBride}, \bibinfo{person}{Michelle~E.
  Minitti}, \bibinfo{person}{Irina Pradler}, {and} \bibinfo{person}{Nicholas~I.
  Boyd}.} \bibinfo{year}{2016}\natexlab{}.
\newblock \showarticletitle{Deconvolution of distinct lithology chemistry
  through oversampling with the Mars Science Laboratory Alpha Particle X-Ray
  Spectrometer}.
\newblock \bibinfo{journal}{\emph{X-Ray Spectrometry}} \bibinfo{volume}{45},
  \bibinfo{number}{3} (\bibinfo{year}{2016}), \bibinfo{pages}{155--161}.
\newblock
\urldef\tempurl%
\url{https://doi.org/10.1002/xrs.2681}
\showDOI{\tempurl}
\showeprint{https://onlinelibrary.wiley.com/doi/pdf/10.1002/xrs.2681}


\bibitem[\protect\citeauthoryear{Wiens, Maurice, Barraclough, Saccoccio,
  Barkley, Bell, Bender, Bernardin, Blaney, Blank, Bouy{\'e}, Bridges, Bultman,
  Ca{\"i}s, Clanton, Clark, Clegg, Cousin, Cremers, Cros, DeFlores, Delapp,
  Dingler, D'Uston, Darby~Dyar, Elliott, Enemark, Fabre, Flores, Forni,
  Gasnault, Hale, Hays, Herkenhoff, Kan, Kirkland, Kouach, Landis, Langevin,
  Lanza, LaRocca, Lasue, Latino, Limonadi, Lindensmith, Little, Mangold,
  Manhes, Mauchien, McKay, Miller, Mooney, Morris, Morrison, Nelson, Newsom,
  Ollila, Ott, Pares, Perez, Poitrasson, Provost, Reiter, Roberts, Romero,
  Sautter, Salazar, Simmonds, Stiglich, Storms, Striebig, Thocaven, Trujillo,
  Ulibarri, Vaniman, Warner, Waterbury, Whitaker, Witt, and Wong-Swanson}{Wiens
  et~al\mbox{.}}{2012}]%
        {Wiens2012}
\bibfield{author}{\bibinfo{person}{Roger~C. Wiens}, \bibinfo{person}{Sylvestre
  Maurice}, \bibinfo{person}{Bruce Barraclough}, \bibinfo{person}{Muriel
  Saccoccio}, \bibinfo{person}{Walter~C. Barkley}, \bibinfo{person}{James~F.
  Bell}, \bibinfo{person}{Steve Bender}, \bibinfo{person}{John Bernardin},
  \bibinfo{person}{Diana Blaney}, \bibinfo{person}{Jennifer Blank},
  \bibinfo{person}{Marc Bouy{\'e}}, \bibinfo{person}{Nathan Bridges},
  \bibinfo{person}{Nathan Bultman}, \bibinfo{person}{Phillippe Ca{\"i}s},
  \bibinfo{person}{Robert~C. Clanton}, \bibinfo{person}{Benton Clark},
  \bibinfo{person}{Samuel Clegg}, \bibinfo{person}{Agnes Cousin},
  \bibinfo{person}{David Cremers}, \bibinfo{person}{Alain Cros},
  \bibinfo{person}{Lauren DeFlores}, \bibinfo{person}{Dorothea Delapp},
  \bibinfo{person}{Robert Dingler}, \bibinfo{person}{Claude D'Uston},
  \bibinfo{person}{M. Darby~Dyar}, \bibinfo{person}{Tom Elliott},
  \bibinfo{person}{Don Enemark}, \bibinfo{person}{Cecile Fabre},
  \bibinfo{person}{Mike Flores}, \bibinfo{person}{Olivier Forni},
  \bibinfo{person}{Olivier Gasnault}, \bibinfo{person}{Thomas Hale},
  \bibinfo{person}{Charles Hays}, \bibinfo{person}{Ken Herkenhoff},
  \bibinfo{person}{Ed Kan}, \bibinfo{person}{Laurel Kirkland},
  \bibinfo{person}{Driss Kouach}, \bibinfo{person}{David Landis},
  \bibinfo{person}{Yves Langevin}, \bibinfo{person}{Nina Lanza},
  \bibinfo{person}{Frank LaRocca}, \bibinfo{person}{Jeremie Lasue},
  \bibinfo{person}{Joseph Latino}, \bibinfo{person}{Daniel Limonadi},
  \bibinfo{person}{Chris Lindensmith}, \bibinfo{person}{Cynthia Little},
  \bibinfo{person}{Nicolas Mangold}, \bibinfo{person}{Gerard Manhes},
  \bibinfo{person}{Patrick Mauchien}, \bibinfo{person}{Christopher McKay},
  \bibinfo{person}{Ed Miller}, \bibinfo{person}{Joe Mooney},
  \bibinfo{person}{Richard~V. Morris}, \bibinfo{person}{Leland Morrison},
  \bibinfo{person}{Tony Nelson}, \bibinfo{person}{Horton Newsom},
  \bibinfo{person}{Ann Ollila}, \bibinfo{person}{Melanie Ott},
  \bibinfo{person}{Laurent Pares}, \bibinfo{person}{Ren{\'e} Perez},
  \bibinfo{person}{Franck Poitrasson}, \bibinfo{person}{Cheryl Provost},
  \bibinfo{person}{Joseph~W. Reiter}, \bibinfo{person}{Tom Roberts},
  \bibinfo{person}{Frank Romero}, \bibinfo{person}{Violaine Sautter},
  \bibinfo{person}{Steven Salazar}, \bibinfo{person}{John~J. Simmonds},
  \bibinfo{person}{Ralph Stiglich}, \bibinfo{person}{Steven Storms},
  \bibinfo{person}{Nicolas Striebig}, \bibinfo{person}{Jean-Jacques Thocaven},
  \bibinfo{person}{Tanner Trujillo}, \bibinfo{person}{Mike Ulibarri},
  \bibinfo{person}{David Vaniman}, \bibinfo{person}{Noah Warner},
  \bibinfo{person}{Rob Waterbury}, \bibinfo{person}{Robert Whitaker},
  \bibinfo{person}{James Witt}, {and} \bibinfo{person}{Belinda Wong-Swanson}.}
  \bibinfo{year}{2012}\natexlab{}.
\newblock \showarticletitle{The ChemCam Instrument Suite on the Mars Science
  Laboratory (MSL) Rover: Body Unit and Combined System Tests}.
\newblock \bibinfo{journal}{\emph{Space Science Reviews}}
  \bibinfo{volume}{170}, \bibinfo{number}{1} (\bibinfo{date}{01 Sep}
  \bibinfo{year}{2012}), \bibinfo{pages}{167--227}.
\newblock
\showISSN{1572-9672}
\urldef\tempurl%
\url{https://doi.org/10.1007/s11214-012-9902-4}
\showDOI{\tempurl}


\bibitem[\protect\citeauthoryear{{Wikipedia contributors}}{{Wikipedia
  contributors}}{2021}]%
        {SpectralDensity}
\bibfield{author}{\bibinfo{person}{{Wikipedia contributors}}.}
  \bibinfo{year}{2021}\natexlab{}.
\newblock \bibinfo{title}{Spectral density --- {Wikipedia}{,} The Free
  Encyclopedia}.
\newblock
\newblock
\urldef\tempurl%
\url{https://en.wikipedia.org/w/index.php?title=Spectral_density&oldid=998058609}
\showURL{%
\tempurl}
\newblock
\shownote{[Online; accessed 12-February-2021].}


\bibitem[\protect\citeauthoryear{Ye, Hermann, Bhat, and Yildirim}{Ye
  et~al\mbox{.}}{2019}]%
        {Repo}
\bibfield{author}{\bibinfo{person}{Constance Ye}, \bibinfo{person}{Lukas
  Hermann}, \bibinfo{person}{Shravya Bhat}, {and} \bibinfo{person}{Nur
  Yildirim}.} \bibinfo{year}{2019}\natexlab{}.
\newblock \bibinfo{title}{PIXLISE-C Source Code}.
\newblock
\newblock
\urldef\tempurl%
\url{https://github.com/cmudig/pixlise-c}
\showURL{%
\tempurl}


\end{thebibliography}

\end{document}